А. А. Крижановский
Санкт-Петербургский институт информатики и автоматизации РАН
aka at iias.spb.su

# Оценка результатов поиска семантически близких слов в Википедии: Information Content и адаптированный HITS алгоритм[1]

# Evaluation experiments on related terms search in Wikipedia: Information Content and Adapted HITS[2]

*Аннотация: Классификация метрик и алгоритмов поиска семантически близких слов в тезаурусах WordNet, Роже и энциклопедии Википедия расширена адаптированным HITS алгоритмом. С помощью экспериментов в Википедии оценены метрика Резника (Information Content), адаптированная к Википедии, и адаптированный алгоритм HITS. Предложен ресурс для оценки семантической близости русских слов.*

*Abstract: The classification of metrics and algorithms search for related terms via WordNet, Roget's Thesaurus, and Wikipedia was extended to include adapted HITS algorithm. Evaluation experiments on Information Content and adapted HITS algorithm are described. The test collection of Russian word pairs with human-assigned similarity judgments is proposed.*

## ВВЕДЕНИЕ

Под семантически близкими словами (СБС) подразумеваются слова близкие по значению, встречающиеся в одном контексте. Это могут быть синонимы (*чертог*, *дворец*), антонимы (*запутать*, *распутать*) и др. Во многих задачах умение составить список СБС, либо сравнить слова и вычислить – какие слова ближе по значению, оказывается востребованным.

Во-первых, это так называемый «поиск по смыслу», при котором пользователь вводит слово *мобильник*, но видит страницы, содержащие другие слова, например *мобильный телефон*, *сотовый* и др. Поисковая система расширила или переформулировала запрос с помощью СБС [Braslavskiy2004], [Ding2005], [Shi2005].

Во-вторых, запросно-ответные системы на этапе обработки вопроса пытаются вычислить, к какой области относится вопрос пользователя, пытаются найти похожие вопросы в базе данных. Поиск вопросов основан в том числе и на использовании списков СБС.

В-третьих, для выбора одного из значений многозначного слова [Resnik2000], [Yarowsky1995] (например, слово *граф* может обозначать либо титул, либо математический объект) используют СБС.

В-четвёртых, есть интерес к автоматическому созданию специальных словарей–тезаурусов на основе СБС [Kashyap2005]. Прелесть таких тезаурусов в том, что они строятся по тексту и могут наглядно, в виде картинки, предъявить ключевые понятия, найденные в тексте, и то, как они связаны.

В-пятых, трудоёмкая задача составления словарей синонимов (и не только синонимов) требует кропотливой работы лексикографов. Своевременную помощь оказывают поисковые алгоритмы, предлагающие списки близких по значению слов для последующего вдумчивого разбора лингвистом.

Количество научных работ, посвящённых Википедии, стремительно растёт.[3]

---

1 Статья доступна по адресу: http://arxiv.org/abs/0710.0169
2 Short version of the paper to be published in Proceedings of the Wiki-Conference 2007, Russia, St. Petersburg, October 27-28. http://ru.wikipedia.org/wiki/Википедия:Вики-конференция_2007/English
3 См. http://en.wikipedia.org/wiki/Wikipedia:Wikipedia_in_academic_studies

Осветим одну из граней этого феномена, а именно: корпус текстов Википедии[4] обладает особой привлекательностью для поисковых алгоритмов. Вики занимает нишу между, с одной стороны, размеченными корпусами текстов, а с другой – интернет-страницами (где нет никаких надёжных подсказок для алгоритмов, кроме гиперссылок и частоты слов). Перечислим «изюминки» вики-текстов с точки зрения машинной обработки:

- заголовок, максимально точно соответствующий теме статьи. Это выгодно отличает вики от других литературных жанров. Например, броский заголовок газетной статьи «Танцуют все!», рассказывающей однако о Википедии, усложнит жизнь поисковику, учитывающему слова из заголовка;
- первый абзац, обычно дающий краткое описание термина, может содержать основные ключевые слова;
- наличие внутренних ссылок на статьи по данной теме; специальный раздел ссылок «*Смотри также*»;
- специальный формат для ссылок на статью о том же термине на другом языке (интервики);
- категории, классифицирующие документы по их тематической принадлежности.

Достоинством Википедии, как корпуса в целом, является большое количество текстов (больше 200 тыс. на русском, больше двух млн. на английском) и доступность дампов[5] энциклопедии.

Системы поиска семантически близких слов в Википедии помогут пользователям, во-первых, находить энциклопедические статьи, близкие по тематике к заданным, что позволит более глубоко изучить исследуемое понятие, а во-вторых, помогут в указании недостающих ссылок между связанными по смыслу статьями.[6]

Далее в статье идёт теоретическая часть, в которой перечислены алгоритмы, применяемые для поиска СБС; рассмотрена мера Резника (Information Content) и её адаптация к таксономии категорий Википедии, описана тестовая коллекция 353-TC. В практической части сравниваются результаты работы AHITS алгоритма с другими на основе данных тезаурусов WordNet, Роже и энциклопедии Википедии.

**АЛГОРИТМЫ ПОИСКА СБС**

Поиск семантически близких слов связан с теорией графов, а именно с анализом веб-ссылок (англ. *web link analysis*) и поиском на основе данных тезауруса [Leontyeva2006], [RuizCasado2005]. Поиск СБС с помощью анализа вес-ссылок основан на следующей предпосылке: *отдельной вершине графа соответствует одна интернет-страница*. При этом отдельной интернет-странице может соответствовать либо понятие[7], либо словоформа[8]. Принятие этой предпосылки позволяет перейти к задаче поиска *похожих интернет страниц*, связанную с задачей вычисления меры сходства между вершинами графа.

---

4   Особенности и потенциал Викисловаря для машинной обработки достойны отдельной статьи.
5   Дамп – это слепок всех данных Википедии в какой-то момент времени. Его можно скачать, и установить Википедию на локальный компьютер.
6   См. http://ru.wikipedia.org/wiki/Википедия:Проект:Связность Следует отметить, что данный алгоритм AHITS не позволит решить проблему страниц-сирот, поскольку для поиска анализирует гиперссылки. Для обработки страниц без ссылок (или с малым их числом) нужен алгоритм, учитывающий частотность слов в корпусе, например, по схеме TF-IDF, см. алгоритм ESA [Gabrilovich2007]. В ESA *концепт* – это название энциклопедической статьи Википедии (ВП). На вход подаются два текста. По ним строятся два вектора из концептов ВП. Для сравнения текстов сравнивают два вектора, например, с помощью косинусного коэффициента.
7   Так в Википедии: странице энциклопедии соответствует некоторое понятие, которое раскрывается в данной энциклопедической статье.
8   Так в Викисловаре: страница словаря описывает одну словоформу, которая может содержать несколько значений.

Для поиска похожих текстовых документов, поиск СБС, вычисление меры сходства между вершинами графа могут использоваться такие алгоритмы, как: Hypertext Induced Topic Selection (HITS) [Kleinberg1999], PageRank [Brin1998], [Fortunato2005] (и его модификация Green [Ollivier2007]), ArcRank [Berry2003], ESA [Gabrilovich2007], алгоритм извлечения синонимов из толкового словаря [Berry2003], алгоритм извлечения контекстно связанных слов [Karypis1999], [Pantel2000] и др.

Алгоритм HITS был адаптирован к поиску в корпусах с гиперссылками и категориями. Реализация алгоритма строит автоматически упорядоченный список СБС в энциклопедии Википедия [Bellomi2005], [Holloway2005], [Ponzetto2006], [Rosenzweig2006], [RuizCasado2005], [Strube2006], [Volkel2006], [Voss2006].

## МЕТРИКА РЕЗНИКА И КАТЕГОРИИ ВИКИПЕДИИ

Учёный Резник [Resnik95] предложил считать, что два слова тем более похожи, чем более информативен концепт (*Information Content*), к которому соотнесены оба слова, то есть чем ниже в таксономии находится общий верхний концепт (синсет в WordNet).[9] Для категорий Википедии *Лётчики* и *Самолёты* ближайшим общим концептом будет *Авиация* (рис. 1).

При построении вероятностной функции *P(C)*, потребуем, чтобы вероятность концепта *C* не уменьшалась при движении вверх по иерархии: $res(c_1, c_2) = max_{C \in S(c_1, c_2)}[-\log(P(C))]$. Тогда более абстрактные концепты будут менее информативны. Резник предложил оценивать вероятность через частоту синонимов концепта в корпусе таким образом:

$$P(C) = \frac{freq(C)}{N}, \quad freq(C) = \sum_{n \in words(C)} count(n)$$, где *words(C)* – это существительные[10], имеющие значение *C*; при этом *N* – общее число существительных в корпусе. Пусть также сходство двух концептов равно нулю, если ближайшим общим концептом является корневой элемент категории.

В работе [Strube2006] метрика Резника *res* была адаптирована к Википедии и информативность категории *P(C)* вычислялась как функция от числа гипонимов категорий, а не статистически:

$$res_{hypo}(c_1, c_2) = 1 - \frac{\log(hypo(lcs_{c_1, c_2}) + 1)}{\log(C)},$$

где *lcs* — ближайший общий родитель концептов *c₁* и *c₂* (от англ. *least common subsumer*); *hypo* — число гипонимов[11] этого родителя; *C* — общее число концептов в иерархии. То есть вместо того, чтобы считать частотность терминов в Википедии (как в оригинальной формуле Резника), Струбе предложил подсчитать число гипонимов. Возможно, это одна из причин, почему мера *res hypo* показала в экспериментах [Strube2006] относительно слабый результат.

На рис. 1 видно, как уменьшается число подкатегорий и статей при спуске по иерархии вниз (hypo — первое число в скобках). Информативность категории *res hypo* при этом увеличивается (второе число). Таким образом, можно вычислить семантическую близость слов. Например, для *дирижабля* и *самолёта* семантическое

---

9 Заметим, что в ВП у слова обычно несколько категорий, то есть может быть несколько ближайших общих категорий.

10 В экспериментах Резник оценивал сходство существительных, учитывал отношение WordNet *IS-A* (гипонимия).

11 Гипонимы категории *K* в Википедии – это все подкатегории *K*, а также все статьи, принадлежащие этим подкатегориям и категории *K*.

сходство равно 0.575, поскольку ближайшим общим концептом будет концепт «*Воздушные суда*».

## 353 ПАРЫ АНГЛИЙСКИХ СЛОВ ДЛЯ ОЦЕНКИ

Для оценки метрик и алгоритмов, вычисляющих близость значений слов, используют тестовый набор (англ. *Test Collection*) из 353 пар английских слов, предложенный в работе [Finkelstein02] (далее 353-ТС).[12]

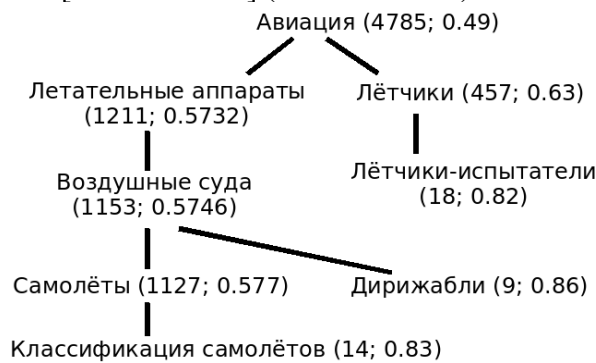

**Рис. 1. Часть иерархии категорий Википедии с указанием числа гипонимов и информативности категории (первое и второе число в скобках)**

Респонденты присвоили значения от 0 до 10 семантической близости парам слов, где 0 указывает на то, что слова совершенно не связаны, 10 – слова почти полные синонимы. В оценке пар слов участвовало 13 человек, обработавших 153 слова, и 16 человек оценивших 200 слов.

Критика данного тестового набора, приведённая в работе [Jarmasz03], заключается в том, что: не приведена методология составления списка, респондентам сложнее давать оценку от 0 до 10, чем на более привычной шкале от 0 до 4. Достоинство данного тестового набора в том, что он:
- превосходит другие тестовые наборы по размеру[13];
- позволяет оценивать семантическую близость, а не только семантическое сходство[14].

## АДАПТИРОВАННЫЙ КОЭФФИЦИЕНТ СПИРМЕНА

Для численной оценки степени сходства эталонного списка и автоматически построенного списка (семантически близких слов) адаптирован коэффициент Спирмена (англ. *Spearman's footrule*). Модификация позволяет сравнивать ранжирование элементов в списках разной длины. Итак, для исходного слова даны: эталонный список *A*, построенный экспертом, и автоматически построенный список *B*. Предлагается добавить в конец списка *B* элементы *A*, в нём отсутствующие. Каждому элементу списка назначается ранг (порядковый номер) от 1 до *N*. Далее применяется формула, где сравниваются положения в списках общих элементов, то есть вычисляется сумма модулей расстояний между *i*-ми элементами набора, *S* – число общих элементов:

---

12 Данные доступны: http://www.cs.technion.ac.il/~gabr/resources/data/wordsim353/wordsim353.html
13 Тесты на синонимию: 80 вопросов теста TOEFL, 50 вопросов ESL [Turney2001] и 300 вопросов Reader's Digest Word Power Game [Jarmasz03].
14 Различают понятия *related terms* (семантически связанные, близкие по значению слова) и *similar terms* – семантически сходные, сходные по значению слова (синонимы). Таким образом, понятие *semantic relatedness* шире, чем *semantic similarity*, так как сюда включаются (кроме синонимии) ещё и отношения меронимии, антонимии и др. [Gabrilovich2007]. AHITS алгоритм позволяет находить семантически близкие слова (*semantic relatedness*).

$$F^S(s_1, s_2) = \sum_{i=1}^{S} (s_1(i) - s_2(i)) .$$

Коэффициент Спирмена позволяет сравнивать с эталонным списком ранжирование одного и того же набора слов AHITS алгоритмом при разных входных параметрах (размер корневого набора, максимально допустимый вес кластера $C_{max}$ и др.).

## ЭКСПЕРИМЕНТЫ[15]

В предыдущих работах [Krizhanovsky2006a], [Krizhanovsky2006b] описан адаптированный HITS (далее AHITS) алгоритм, представлены эксперименты по поиску синонимов в английской и русской версии Википедии с помощью AHITS алгоритма и описана сессия поиска синонимов в программе *Synarcher*. Далее в данной работе описаны результаты и особенности вычисления метрики Information Content ($res_{hypo}$) и результаты работы алгоритма AHITS.

*Метрика Резника, адаптированная к Википедии*

Эксперименты по вычислению метрики $res_{hypo}$ в википедиях на английском, simple[16] и русском языках показали, что есть некоторые особенности, определяемые структурой Википедии: (i) в графе категорий есть циклы; (ii) стоит задача выбора корневой категории.

*Циклы*

Метрика $res_{hypo}$ рассчитана на дерево без циклов, но это не так в Википедии. Для вычисления метрики по данным дампа Википедии написана хранимая MySQL процедура. При вычислении получено 16'777'215 гипонимов у категории *Main_topic_classifications*, что не верно, поскольку *превышает* суммарное число категорий и статей 3'978'376 (а категорий всего 244'618) в английской Википедии (дамп от 27 мая 2007). Вероятно, это обусловлено наличием циклов в структуре категорий. Найдено 526 циклов[17] в дампе английской Википедии и 34 цикла в русской.[18]

*Корневая категория*

Дело в том, что некоторые из категорий первого уровня[19] выполняют вспомогательную функцию, например, *Категория:Википедия* в русской или *Category:Wikipedia_categories* в английской. Они содержат служебные категории, которые, вероятно, если их учитывать, ухудшат работу метрики $res_{hypo}$. Чтобы проверить эту гипотезу, была вычислена метрика $res_{hypo}$ для каждой категории, при этом эксперимент проведён для нескольких корневых категорией (столбец *Корневая категория* в табл. 1).

Время предварительного вычисления метрики $res_{hypo}$ для каждой категории указано указано в графе *Оффлайн* в часах. Время поиска общего концепта с наименьшим значением $res_{hypo}$ для всех 353 пар слов указано в графе *Онлайн*.

Данный эксперимент не решил поставленный вопрос, поскольку ухудшение значения корреляции (0.33 по сравнению с 0.36-0.37) можно объяснить не только тем, что *Category:Categories* включает *Category:Wikipedia_categories*, а также тем, что изменилось число сравниваемых пар (графа *Пропущено* в табл. 1). Эта графа указывает на число пар слов в 353-TC, для которых либо не было найдено общих концептов, либо общим концептом является только корневая категория.

---

[15] Эксперименты можно повторить с помощью программы Synarcher версии 0.12.4, см. *Release Notes* в программе, http://synarcher.sourceforge.net
[16] Простая Английская Википедия, см. http://simple.wikipedia.org
[17] См. список циклов на стр. http://en.wikipedia.org/wiki/User:AKA_MBG/Cycles
[18] См. http://ru.wikipedia.org/wiki/Википедия:Проект:Систематизация_категорий/Совместная_работа
[19] Назовём для удобства корневую категорию Википедии категорией нулевого уровня, то есть категория в таксономии находится на глубине ноль. В английской Википедии это категория *Categories*.

Для чистого эксперимента нужно будет учесть только те пары, которые найдены во всех трёх случаях. Хотя вывод о том, что использование категории нулевого уровня (по сравнению с категориями первого уровня) увеличивает покрытие тестового набора почти в два раза, можно сделать уже сейчас.

**Табл. 1. Корреляции метрики *res $_{hypo}$*. с оценками респондентов 353-TC в зависимости от выбора корневой категории на основе данных английской Википедии (версия от 27 мая 2007)**

| Корневая категория | Глубина | Корреляция | Пропущено | Оффлайн, ч | Онлайн, сек |
|---|---|---|---|---|---|
| Main_topic_classifications | 1 | 0.36 | 149 | 27.5 | 24 |
| Fundamental | 1 | 0.37 | 150 | 24.1 | 30 |
| Categories | 0 | 0.33 | 15 | 30.6 | 21.2 |

*Оценка метрики Резника и алгоритма AHITS*

Коэффициент Спирмена получилось приспособить для оценки корреляции результатов работы алгоритма AHITS и тестового набора 353-TC. Для этого берётся пара английских слов из 353-TC. Запускается дважды алгоритм AHITS и для двух слов получены два списка, например, из 1000 семантически близких слов заданному. Далее эти два списка подставляются в формулу вычисления адаптированного коэффициента Спирмена (графа *Spearm. footrule* в табл. 2). Либо эти два списка пересекаются, чтобы получить число общих слов в этих списках (графа *число общих слов* в табл. 2). Было неожиданно получить на английской Википедии (строка *English 20070527* в табл. 2) то, что число общих слов лучше коррелирует с 353-TC, чем адаптированный коэффициент Спирмена.

**Табл. 2. Корреляция результатов вычисления метрики *res $_{hypo}$* и работы алгоритма *AHITS* на основе данных Википедии с оценками семантической близости слов респондентов**

| Данные | Корреляция | | | Всего | | | | |
|---|---|---|---|---|---|---|---|---|
| | res $_{hypo}$ | AHITS | | res $_{hypo}$ | | | AHITS | |
| | | Spearm. footrule | Число общих слов | Пропу-щено | Время | | Пропу-щено | Время онлайн ч |
| | | | | | оффлайн, ч | онлайн, сек | | |
| Simple 20070811 | – | **0.4** | 0.33 | – | – | – | 203 | 0.08 |
| Simple 20070909 | 0.37 | 0.15 | 0.31 | 155 | – | 19 | 192 | 0.06 |
| English 20060502 | – | **0.4** | **0.39** | – | – | – | 157 | 17.01 |
| *(AHITS low load) English 20070527* | – | *0.14* | *0.16* | – | – | – | *90* | *2.31* |
| (*AHITS* high load) English 20070527 | 0.33-0.36[20] | 0.16 | **0.38** | 15-149[21] | 27.5-30.6 | 21-24 | **29**[22] | 49.82 |

---

20 См. столбец *Корреляция* в табл. 1.
21 См. столбец *Пропущено* в табл. 1.
22 Требуется провести дополнительное исследование, чтобы выяснить, почему для 29 пар слов не было найдено общих слов.

Таким образом, оценка корреляции результатов поиска СБС с тестовым набором 353-TC показала, что алгоритм AHITS даёт несколько лучший результат (0.38-0.39), чем адаптированная метрика Резника (0.33-0.36) на данных английской Википедии. В экспериментах с Simple Википедией получен значительный разброс значений корреляции для AHITS: от 0.15 до 0.4 (столбец *Spearm. footrule* в табл. 2).

Достоинства и недостатки энциклопедии Simple (SW) заключаются в том. что она на два порядка меньше английской (20 тыс. страниц против двух миллионов на сентябрь 2007 г). Плюс SW в том, что на ней удобно отлаживать алгоритмы, а недостаток в том, что меньшее покрытие тем даёт худший результат. Например, в SW оказалось пропущено 155 пар слов, а в английской от 15 до 149.

Для алгоритма AHITS не указано время оффлайн, поскольку предобработка отсутствует. Чтобы показать важность параметров алгоритма AHITS, в табл. 2 добавлена строки:
- *AHITS low load* соответствует эксперименту с такими параметрами: корневой набор – 3 вершины, инкремент – 1 вершина, число искомых СБС – 10;
- *AHITS high load*: корневой набор – 200 вершин, инкремент – 17 вершин, число искомых СБС – 1000;

В таблице видно, что режим *AHITS high load* потребовал в 20 раз больше времени для вычислений, корреляция с данными респондентов повысилась с 0.16 до 0.38. Возможно, эксперимент не очень честный и нужно было бы установить параметр «*число искомых СБС*» одинаковым в обоих случаях, поскольку он непосредственно влияет на значение корреляции. Что останется без изменений, так это время поиска.

*Результаты и классификация*

Центральное место в данной работе занимает таблица с оценкой работы алгоритмов и метрик. Столбцы *AHITS* и *res hypo* подытоживают эксперименты, представленные в табл. 2. Данные для других метрик и алгоритмов в основном взяты из работы [Strube2006], в ней также описаны метрики *jaccard*, *text*, *res hypo*. Использованы экспериментальные данные таких работ, как [Jarmasz03] (*jarmasz*), [Finkelstein02] (поисковик IntelliZap и алгоритм *LSA*), [Gabrilovich2007] (алгоритм *ESA*). Представление об остальных метриках можно получить из работ: [WuPalmer94] метрика *wup*, ([Fellbaum1998], стр. 265-283) метрика *lch*, [Resnik95] метрика *res*, [Banerjee02] метрика *lesk*.

Классификация метрик и алгоритмов поиска СБС, предложенная в [Strube2006], расширена (1) адаптированным HITS алгоритмом, основанном на анализе веб-ссылок, и (2) явным указанием отдельной группы методов, полагающихся на частотность слов в корпусе. Таким образом, предложена следующая классификация (табл. 3) метрик и алгоритмов поиска СБС, основанных на учёте (i) расстояния в таксономии, (ii) анализа веб-ссылок, (iii) частотности слов в корпусе, (iv) совпадения (перекрытия) текстов. Отметим, что метрика Резника *res* учитывает одновременно и частотность слов, и свойства (не расстояние) концептов в таксономии.

Табл. 3 содержит значения корреляции тестовой коллекцией 353-TC и результатов, полученных с помощью указанных метрик и алгоритмов. Получены лучшие результаты при поиске с учётом:
- *расстояния в таксономии* – 0.48, метрика *lch* ([Fellbaum1998], стр. 265-283) для английской Википедии;
- *частотности слов в корпусе* – 0.75, алгоритм *ESA* [Gabrilovich2007] для английской Википедии;
- *перекрытия текстов – 0.21,* метрика *lesk* [Banerjee02] для тезауруса WordNet.

Вне рассмотрения оставлен алгоритм Green [Ollivier2007] (поиск в Википедии), поскольку нет данных о его тестировании с помощью коллекции 353-TC.

**Табл. 3. Классификация алгоритмов и корреляция результатов с данными респондентов
(на данных тестового набора 353-TC, без пропусков)**

| Набор данных | Расстояние в таксономии | | | | Анализ ссылок | Частотность слов в корпусе | | | | Перекрытие текстов | |
|---|---|---|---|---|---|---|---|---|---|---|---|
| | wup | lch | res $_{hypo}$ | jarmasz | AHITS | jaccard | res | LSA | ESA | lesk | text |
| WordNet | 0.3 | 0.34 | – | – | – | – | 0.34 | | – | 0.21 | – |
| Wiki-pedia[23] | 0.47 | **0.48** | *0.33-0.36 0.37*[24] | – | *0.38-0.39*[25] | – | –[26] | | **0.75** | 0.2 | 0.19 |
| Simple Wikipedia | – | – | *0.37* | – | *0.31-0.33* | – | – | | – | – | – |
| Другие | – | – | – | Тезаурус Роже 0.539[27] | – | Google 0.18 | – | IntelliZap 0.56 | | – | – |

### ЗАКЛЮЧЕНИЕ И ТЕСТОВЫЙ НАБОР РУССКИХ СЛОВ

П. Л. Капица писал: «...теория — это хорошая вещь, но правильный эксперимент остаётся навсегда»[28]. Однако, чтобы провести эксперимент и оценить результаты поиска близких по значению слов, нужен тестовый набор (эталон), который создан людьми вручную, а не автоматически.

Для английского языка такой набор есть – это 353 пары слов, в оценке которых участвовало два десятка людей. Табл. 3 показывает, что уже более десяти метрик и алгоритмов можно сравнить с помощью этих данных. Именно этот набор использовался и для оценки работы программы *Synarcher*, реализующей адаптированный HITS алгоритм, в английской и английской простой Википедиях.

Было бы интересно оценить работу алгоритмов в русской Википедии. Предлагаю проставить оценки в тестовом наборе из русских слов и приглашаю на страницу проекта[29].

### БЛАГОДАРНОСТИ

# Список источников литературы

---

23 Английская Википедия, см. http://en.wikipedia.org
24 0.33-0.36 см. в табл. 2, 0.37 взято из [Strube2006].
25 См. табл. 2.
26 Сомнения по поводу того, чтобы считать эквивалентными метрики *res* [Resnik95] и *res* $_{hypo}$ [Strube2006] изложены на стр. 3.
27 0.539, см. [Jarmasz03], стр. 4. Значение 0.55 в работе [Gabrilovich2007] - это, вероятно, опечатка.
28 Капица П. Л. Эксперимент, теория, практика. М., 1974. 288 с. http://whinger.narod.ru/ocr/
29 См. http://ru.wikipedia.org/wiki/Участник:AKA_MBG/Wordsim